\documentclass [11pt]{article}
\usepackage {amssymb}
\usepackage {amsmath}
\usepackage {graphicx}
\def\be{\begin{equation}}
\def\ee{\end{equation}}

\begin{document}
\renewcommand{\baselinestretch}{1.2}

\title{Evolutionary tradeoff and equilibrium in an aquatic predator-prey system}         
\author{Laura E. Jones\footnote{lej4@cornell.edu} \\
Stephen P. Ellner\footnote{spe2@cornell.edu}}
 
\maketitle

\large
\begin{center}
\textsc{Cornell University} \\
Department of Ecology and Evolutionary Biology \\
E145 Corson Hall \\
Ithaca, New York 14853-2701
\end{center} 

\pagebreak
\small
\section*{Abstract}
Due to the conventional distinction between {\em ecological} (rapid) and {\em evolutionary} (slow)
timescales, ecological and population models to date have typically ignored the effects of
evolution.  Yet the {\em potential} for rapid evolutionary 
change has been recently established and may be critical to 
understanding how populations adapt to changing environments.
In this paper we examine the relationship between ecological and evolutionary 
dynamics, focusing on a well-studied experimental aquatic predator-prey system 
(Fussmann et al. 2000; Shertzer et al. 2002; Yoshida et al. 2003). 
Major properties of predator-prey cycles in this system are determined by ongoing 
evolutionary dynamics in the prey population.
Under some conditions, however, the populations tend to apparently stable 
steady-state densities. These are the subject of the present paper. We 
examine a previously developed model for the system, to determine how evolution shapes 
properties of the equilibria, in particular the number and identity of coexisting
prey genotypes. We then apply these results to explore how 
evolutionary dynamics can shape the responses of the system to ``management": externally imposed 
alterations in conditions. Specifically, we compare the behavior of 
the system including evolutionary dynamics, with predictions that would be made 
if the potential for rapid evolutionary change is neglected. 
Finally, we posit some simple experiments to 
verify our prediction that evolution can have significant 
qualitative effects on observed population-level responses 
to changing conditions.

\normalsize 
\section{Introduction}

A distinction is often made between \textit{ecological} and
\textit{evolutionary} time scales, the former referring to relatively
rapid changes in the distribution and abundance of species, the latter
to more gradual changes in the properties of those species as a 
result of natural selection. This distinction is tacitly embedded in much
of the established theory in ecology and evolution. Ecological models
conventionally assume constant parameters to characterize species and their
interactions (a recent influential text on community ecology (Morin 1999) does not
even list evolution among the ``factors influencing interactions among species'').
Conversely, many evolutionary and population-genetic models assume 
constant population size on the grounds that (relative to the evolutionary 
time scale) ongoing changes in population size are merely high-frequency ``noise''. 
When population dynamics are considered, only rare extreme events 
(bottlenecks, expansion after colonization, etc.) are assumed to matter. Even
in theories that explicitly combine ecological and evolutionary dynamics, the
separation of time-scales is typically assumed. For example, Khibnik and Kondrashov (1997)
analyze predator-prey coevolution using singular perturbation theory, assuming that 
ecological dynamics are much faster than evolution. The ``canonical equation''
of Adaptive Dynamics (Marrow et al. 1996, Dieckmann et al. 1997) used in many recent 
papers (e.g. Dercole et al. 2003, Le Galliard et al. 2003) goes even further,
and assumes evolutionary changes are so rare that ecological dynamics reach
their new asymptotic steady-state or attractor before another evolutionary step occurs.  

However, recent years have seen an accumulation of evidence that this presumed
separation of time scales is often violated (e.g., Thompson 1998, Hairston et al. 1999, 
Cousyn et al. 2001, Reznick and Ghalambor 2001, Grant and Grant 2002, Yoshida et al. 2003): 
in natural and experimental settings, trait evolution and changes in abundance may occur
on similar time scales. For example, Resnick et al. (1997) observed
significant life history evolution in guppies over periods of 4-11 years in 
response to changes in predation pressure, and estimated that 
evolution occurred in these populations at rates ``up to seven orders of magnitude greater than 
rates inferred from the paleontological record'' (Resnick et al. 1997, p. 1934). 
Ashley et al. (2003) review evidence for rapid evolutionary changes in response
to natural and anthropogenic agents in birds, fishes, mammals, lizards, and plants, with
changes occuring in some cases within as little as a single year or generation.  

In addition to its theoretical implications, the potential for rapid evolutionary 
change is critical to understanding how populations adapt to changing
environments, for example in harvested or endangered populations where rapid
changes in conditions can result from human management or lack thereof (Ashley et 
al 2003, Stockwell et al. 2003, Zimmer 2003).  
A change in extrinsic factors affecting mortality or reproductive success -- such as a change in 
harvesting rate or juvenile survival -- is necessarily a change in the forces of
natural selection, and the response is likely to be modified by adaptation. 
Yet, based on the assumed
separation of time scales (and despite evidence that commercial fishing has
already led to significant evolutionary changes in harvested species), 
plans for species management rarely take into account the likelihood that ecological
changes will evoke an evolutionary response (Stockwell et al. 2003, Ashley et al. 2003). 

In this paper, we continue our studies of the interplay between ecological and evolutionary 
dynamics, focused on an experimental aquatic predator-prey system 
(Fussmann et al. 2000; Shertzer et al. 2002; Yoshida et al. 2003). 
Previous studies (summarized below) have shown that major properties of predator-prey 
oscillations in this system are determined by ongoing evolutionary dynamics in the prey population.
Under other conditions, however, the predator and prey tend to apparently stable 
steady-state densities. These are the subject of the present paper. We study
the previously developed model for our system under steady-state conditions, 
to determine how evolution shapes properties of the steady states. Specifically,
we ask: 
\begin{enumerate}
\item Under what conditions do we expect to find
single versus multiple genotypes maintained by selection, and which ones? 
\item To what extent does evolution affect the responses of the system
to changes in external conditions such as resource availability and 
extrinsic sources of mortality? 
\end{enumerate} 

The idea that evolution can affect ecological dynamics is certainly not
new, indeed a substantial body of theory (reviewed by Abrams 2000) has developed 
over several decades to predict how the dynamics and stability of predator-prey interactions 
can be affected by trait evolution, and how selective harvesting can affect evolution in
fish populations (Law and Grey 1989). However, key assumptions of the 
models have rarely been verified experimentally, 
and the practical difficulties of monitoring long-term changes in population abundances 
and trait distributions in order to test model predictions ``has left a rather unfortunate gap 
between theory and experiment" (Abrams 2000, p. 98). ``Although over 40 years of theory have
addressed how evolutionary processes can affect the ecology of predator prey interactions,
few empirical data have addressed the same issue" (Johnson and Agrawal 2003). 
In the absence of an empirical foundation, there has been a proliferation of speculative models 
and a corresponding proliferation of alternative predictions: ``Evolution can stabilize or 
destabilize interactions $\ldots$ When population cycles exist, adaptation may either increase 
or decrease the amplitude of those cycles" (Abrams 2000). In a general model of 
a 3-species (resource, consumer, predator) food chain, Abrams and Vos (in press) have 
shown that, as a result of adaptive changes in the consumer, an increase in 
consumer mortality rate could result in either an increase or a decrease in the 
resource, consumer (prey), and predator steady state values. 

Our motivation for adding to this literature is to develop specific 
predictions for an experimentally tractable system, where model assumptions and
predictions can be tested with quantitative rigor, allowing direct feedbacks between 
theory and experiment as in our previous studies with this system (Fussmann et al 2000,
Shertzer et al. 2002, Yoshida et al. 2003). Experimental tests of theoretical
predictions are strongest when predictions have been put on record 
\textit{before} the experiments are done (Nelson G. Hairston, Sr., \textit{personal
communication}). We begin by describing the experimental system 
and model. We then analyze the model using approaches from evolutionary game theory -- 
that is, under a given set of experimental conditions we seek to identify 
either single ESS prey genotypes (ESS=evolutionarily stable strategy) that cannot 
be invaded by any alternative genotype, or a non-invasible ESC of coexisting 
genotypes (ESC=evolutionarily stable combination). We then use the results of 
this analysis to predict how population response to changing conditions will be
modified by adaptive changes in the prey. 

The model can produce a wide range of dynamical behaviors depending on parameters. 
A complete analysis of its behavior would be quite lengthy, but also not germane for empirical studies. 
We therefore consider a wide range of values for parameters that are under experimental control, 
but for parameters characterizing the organisms we restrict the analysis to values near 
those estimated to hold in our system. Note that experimental conditions leading to
population cycles require a totally different analysis which will be the subject of
a later paper.

\subsection{Experimental system}
The experimental system is a predator-prey microcosm with rotifers, 
\textit{Brachionus calyciflorus}, and their algal prey, \textit{Chlorella vulgaris}, 
cultured together in nitrogen-limited, continuous flow-through chemostat systems. 
\textit{Brachionus} in the wild are facultatively sexual, but because sexually 
produced eggs wash out of the chemostat before offspring hatch, our rotifer cultures 
have evolved to be entirely parthenogenic (Fussmann \textit{et al.} 2003). The algae 
also reproduce asexually (Pickett-Heaps 1975), so evolutionary change occurs as a result
of changes in the relative frequency of different algal clones. 

Two parameters of the system can be set experimentally: the concentration of limiting nutrient 
[nitrate] in the inflowing culture medium $N_I$, and the dilution rate $\delta$, the 
fraction of chemostat medium that is replaced each day. A simple ordinary differential 
equation model given below is able to capture the experimentally observed qualitative 
behavior of the system: equilibria at low dilution rates, followed by cycling, followed by 
equilibria and then extinction (Fussmann \textit{et al.} 2000) as the dilution rate is increased. However, 
the experimental system exhibited longer-period cycles and unique phase 
relations which did not match the short-period
cycles and classic quarter-cycle predator-prey phase relations predicted by the model. 
Furthermore, the observed cycles showed extended periods where algal densities were high yet 
rotifer numbers remained low, followed by rapid growth in the rotifer populations while algal 
densities remained roughly unchanged. 

A series of models were devised to explain these observations (Shertzer \textit{et al.} 2002), 
each including some biologically plausible mechanism: rotifer self-limitation via reduced egg fitness
when food is scarce; changes in algal nutrient composition as a function of nutrient 
availability; changes in algal physiology due to accumulation of toxins released by rotifers, 
and evolved prey defense against rotifer predation. Only the model including prey evolution was able
to account well for the qualitative properties of the observed cycles. In that model, 
algae exposed to rotifer predation pressure evolved a ``low palatability"  phenotype at some cost to 
their competitive fitness. Subsequent experiments confirmed the existence of an 
evolutionary tradeoff between defense against predation and ``fitness'' (growth rate in absence of 
rotifers): ``grazed'' algae grown under constant rotifer 
predation pressure were smaller, competitively inferior, and constituted inferior food 
(i.e., rotifers grew poorly when fed upon these cells) relative to ``ungrazed'' 
algae grown in rotifer-free environments (Yoshida \textit{et al.}, submitted). Phenotypic
differences between grazed and ungrazed algal lineages were heritable -- 
persisting in subsequent generations grown under common conditions -- demonstrating
that the algal population evolved in response to grazing pressure. Experiments also verified the 
prediction that predator-prey cycles would exhibit very different qualitative properties 
(shorter period, and different phase relations) in the absence of prey evolution (Yoshida 
\textit{et al.} 2003). 

\section{Description of the Model}

\begin{table}
\begin{center}
\small
\centerline{\textbf{Table 1. Model Parameters}}
\begin{tabular}{|p{40pt}|p{160pt}|p{120pt}|p{40pt}|}
\hline
Parameter & Description & Value & Reference \\
\hline
$N_I$ & Concentration of limiting nutrient in supplied medium& $80 \mu$ mol N$/l$ & Set \\
$\delta$ & Chemostat dilution rate & variable ($d$) &  Set  \\
$V$ & Chemostat volume & $0.33 l$ & set  \\
$\chi_c$ & Algal conversion efficiency & variable &  fitted   \\
$\chi_b$ & Rotifer conversion efficiency & variable & fitted   \\
$m$ & Rotifer mortality & $0.055 /d$ &  F2000 \\
$\lambda$ & Rotifer senescence rate & $0.4 /d$ & F2000 \\
$K_c$ & Minimum algal half-saturation & $4.3$ $\mu$ mol N $/l$  & F2000 \\
$K_b$ & Rotifer half-saturation & $0.835 \times 10^9$ algal cells $/l$ &   \\
$\beta_c$ & Maximum algal recruitment rate & $3.3 /d$ & TY  \\
$\omega_c$ & N content in $10^9$ algal cells & $20.0$ $\mu$ mol & F2000  \\
$\epsilon_c$ & Algal assimilation efficiency & $1$ & F2000  \\
$G$ & Rotifer maximum consumption rate& $5.0\times10^{-4}$ $l/d$  & TY \\
$\alpha_1$ & Shape parameter in algal tradeoff  & variable, $\alpha_1 > 0$ & fitted \\
$\alpha_2$ & Scale parameter in algal tradeoff  & variable, $\alpha_2 > 0$ & fitted \\
\hline
\end{tabular}
\end{center}
\textbf{Set}: Adjustable experimental parameter\\
\textbf{F2000}: Fussmann \textit{et al.} 2000\\
\textbf{TY}: Yoshida \textit{et al.}, in prep.
\end{table}

The model is a system of ordinary differential equations describing the predator-prey 
and prey evolutionary dynamics in our microcosms. It is essentially the same as
the model used by Yoshida et al. (2003), with minor simplifications for
the sake of analytic tractability. The possibility for genetic variability and
evolution in the algal prey population is modeled by explicitly representing the algal
population as a finite set of asexually reproducing clones. Each clone is characterized
by its ``food value'' to the rotifers, denoted $p$ (for ``palatability''). Food
value is defined by its effect on the rotifers: $p$ can be thought of as the conditional 
probability that an algal cell is digested rather than ejected, once it has been ingested. 
The model thus consists of three equations for the limiting nutrient and
rotifers, plus $q$ equations for a suite of $q$ algal clones.  In the following equations,
$N$ is nitrogen ($\mu$ mol per liter), $C_i$ represents concentration (per liter) of
the $i^{th}$ algal clone, where 
$i = 1,2, ....,q$; and $R$ and $B$ are \textit{Brachionus}, 
fertile and total population counts in individuals per liter, respectively. Fecund rotifers 
senesce and stop breeding at a rate $\lambda$; all rotifers are subject to fixed mortality $m$.
The parameters $\chi_c, \chi_b$ are conversions between consumption and recruitment
rates (additional model parameters are defined in Table I). 

\begin{eqnarray}
\frac{dN}{dt} &  = & \delta(N_I - N)  - \sum_{i=1}^{q}F_{C,i}(N)C_i  
\label{eq1} \\
\frac{dC_i}{dt} & =  &\chi_c F_{C,i}(N)C_i  - F_{b_i}(C_i)B - \delta C_i 
\label{eq2} \\
\frac{dR}{dt} & = & \chi_b\sum_{i=1}^{q}F_{b_i}(C_i)R - (\delta + m + \lambda)R
\label{eq3} \\
\frac{dB}{dt} & = & \chi_b\sum_{i=1}^{q}F_{b_i}(C_i)R- (\delta + m)B 
\label{eq4} 
\end{eqnarray}

where 
\be
F_{c,i}(N)= \frac{\rho_c N}{K_c(p_i) + N}
\label{eq5}
\ee
and
\be
F_{b_i}(C_i)=\frac{G C_i p_i}{K_b + \sum_{i=1}^{q}C_i p_i}.
\label{eq6} 
\ee
are functional responses describing algal and rotifer consumption rates, respectively, 
and where  $\rho_c = \frac{\omega_c\beta_c}{\epsilon_c}$.

Equation (\ref{eq6}) is derived from the rotifer clearance rate \textbf{G} (the volume of water per
unit time that an individual filters to obtain food), which in this model is a
function of algal food value:
\[
\textbf{G}=\frac{G}{K_b + \sum_{i=1}^{q}C_i p_i}.
\]
That is, algae of lower value result in the rotifers increasing their clearance rate, 
exactly as if the density of food were lower. 
An alternate model in which rotifer clearance rate does not respond to food value 
was also considered, but could not fit the experimental data on population cycles as well. \\

\begin{figure}
\centerline{\includegraphics[width=15.25cm,height=7.62cm]{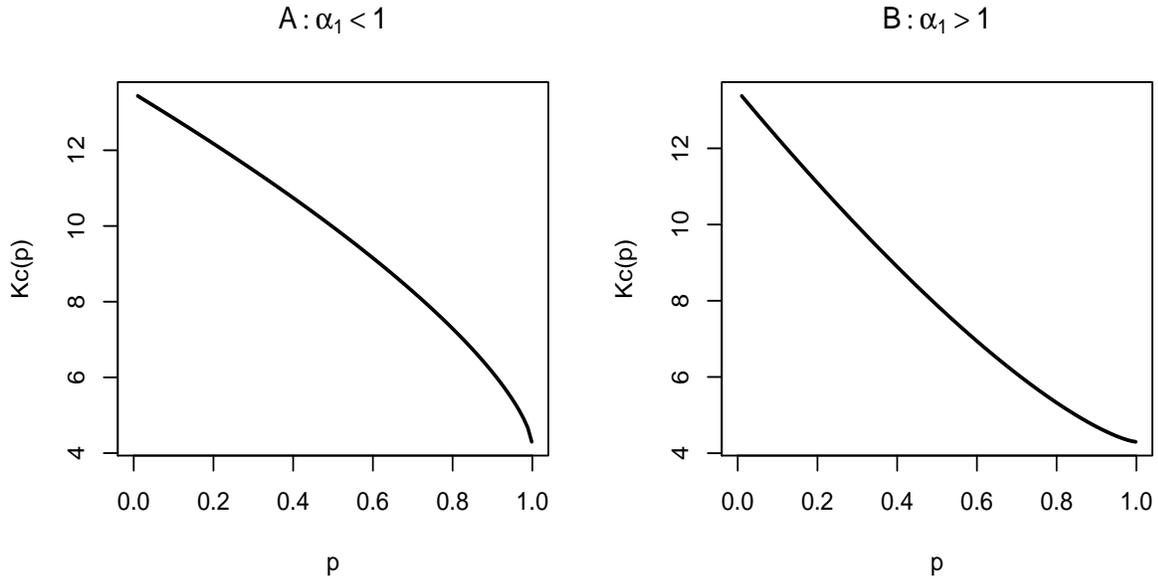}}
\caption{\small Plots of clonal food value, $p$ versus half-saturation, $K_c(p)$ for shape 
parameter $\alpha_1 < 1$ (panel A) and $\alpha_1 > 1$ (panel B).  When the tradeoff curve 
is concave \emph{down} (left panel), it is more likely that surviving clones, either in equilibrium 
or in a cycling regime, represent two extremes in food value. When the curve is concave \emph{up} 
(right panel), intermediate algal types can persist.}
\label{Figure1}
\end{figure}

 Each algal clone $c_i$ is assigned a food value $p_i$ between $1$ (``good") and 0 
(``bad"), with lower $p_i$ resulting in reduced risk of predation. A reduced 
food value comes, however, at the cost of reduced ability to compete for scarce nutrients. 
This relationship is specified by a tradeoff curve [Figure \ref{Figure1}], 
modified from Yoshida {\it et al.} 2003. For simplicity, we use a two-parameter family 
to model how $K_c$ varies between $p_i=0$ and $p_i=1$ :
\[
K_c(p_i) = K_c + \alpha_2(1-p_i)^{\alpha_1}
\]
where $K_c>0$ is the minimum half-saturation value and $\alpha_1, \alpha_2 > 0$
are shape and cost parameters, respectively. 

\begin{figure}
\centerline{\includegraphics[width=15.25cm,height=7.62cm]{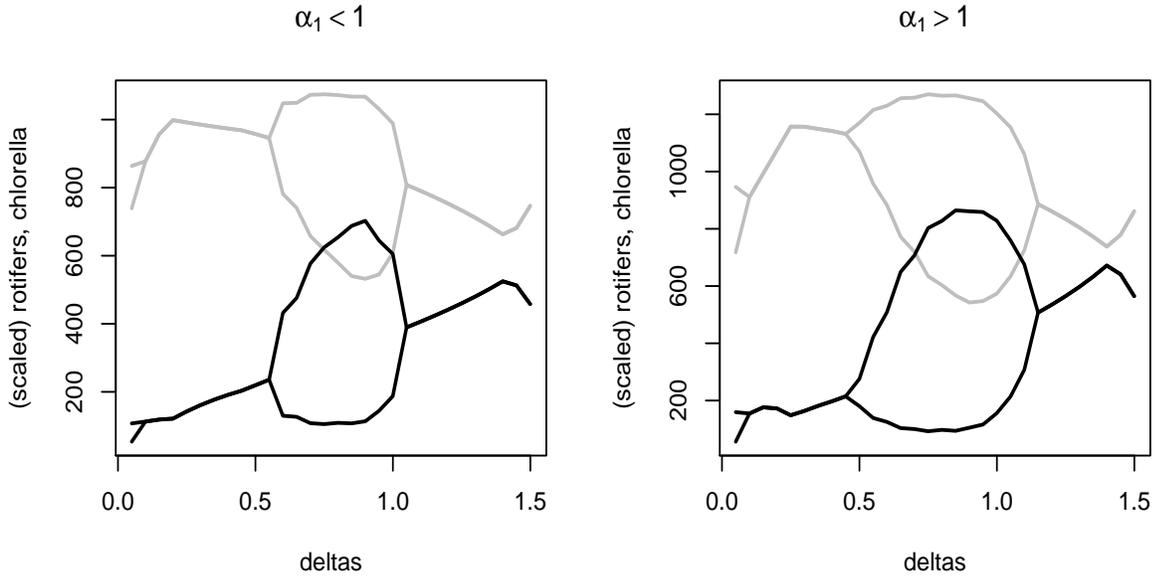}}
\caption{\small Numerical bifurcation diagrams for the clonal model. Population densities of algae (\textit{Chlorella}) 
are shown in grey line; rotifers (\textit{Brachionus}) in black line.  Note the existence of equilibria 
at both low and high dilution rates. Predator-prey cycles occur at intermediate dilution rates.
Parameter sets used in these calculations were obtained by simultaneous fitting of the clonal 
model [(\ref{eq1}) -- (\ref{eq6})] to amplitude, phase, and phase--lag observations 
from an intermediate dilution rate regime (i.e., $\delta=0.69$), and a high dilution rate regime ($\delta=0.96$) 
during which the system was cycling. Left panel, Phase diagram for tradeoff parameter $\alpha_1 < 1$:
$\alpha_1=0.78$, $\alpha_2=9.4$,  and $p_{min}=0.10$.  Right panel, Phase diagram for $\alpha_1 > 1$:
$\alpha_1=1.35$, $\alpha_2=9.6$, and $p_{min}=0.10$. }
\label{Figure2}
\end{figure}

For most parameters we have fairly secure values
based on direct experimental measurements (Table 1). For example,
the parameters defining rotifer feeding rate were estimated by 
allowing known numbers of rotifers to graze on algae at known initial densities, with algal
density measured again after a fixed amount of feeding time (T. Yoshida, \textit{unpublished}). 
For some parameters, however,
we only have indirect estimates obtained by fitting the model to population count
data -- these parameters are described as ``fitted'' in Table 1. Specifically, 
parameters were chosen to match as closely as possible the amplitude and period of predator
and prey cycles, and the phase lag between them, in experimental observations at two
different dilution rates where the system exhibits cycles ($\delta=0.69$ and $\delta=0.96$),
and to match the observed lack of cycles at low and high dilution rates. 
As is often the case, there is a range of parameter values more or less equally consistent
with the data, and we explore how this parameter uncertainty affects our predictions. 
With the best-fitting parameter values, the model ((\ref{eq1})--(\ref{eq6})) 
produces a bifurcation diagram as a function of dilution rate, $\delta$, which approximates the behavior 
our system exhibits in the laboratory (Fussmann et al. 2000, Yoshida \textit{et al.} 2003, Supplementary material). 
``Low-flow'' equilibria exist for dilution rates $\delta \approx 0.05 - 0.50$, followed by cycling at intermediate dilution rates, 
followed by ``high flow'' equilibria at $\delta \approx 1.0 - 1.5$ [Figure \ref{Figure2}]. 

\section{Analysis}
\subsection{Single-clone steady states} 

Assume now one dominant algal clone $C$ with food value $p$. The model (\ref{eq1})--(\ref{eq4}) then reduces to 
the following four equations, after substituting in the appropriate functional 
responses (\ref{eq5})--(\ref{eq6}):
\begin{equation}
\label{eqs7}
\begin{aligned}
\frac{dN}{dt} &  = & \delta(N_I -N)  - \frac{\rho_c NC}{K_c(p) + N}  \\
\frac{dC}{dt} & =  & \chi_c\frac{\rho_c NC}{K_c(p) + N}-\frac{GCpB}{(K_b + Cp)}-\delta C\\
\frac{dR}{dt} & = & \chi_b\frac{GCpR}{K_b + Cp} - (\delta + m + \lambda)R \\
\frac{dB}{dt} & = & \chi_b\frac{GCpR}{K_b + Cp} - (\delta + m)B. 
\end{aligned}
\end{equation}

After some algebra, the steady state for (\ref{eqs7}) is found to be:
\begin{equation}
\label{eqs8}
\begin{aligned}
\bar N & = & \frac{1}{2}[ -\gamma + \sqrt{\gamma^2 + 4N_I K_c(p)}] \\
\bar C &  = & \frac{K_b(\delta + m +\lambda)}{p(\chi_b G - (\delta + m + \lambda))} \\
\bar R & = & \frac{\delta\chi_b(\delta + m)[\chi_c(N_I -\bar N)-\bar C]}{(\delta + m + \lambda)^2} \\
\bar B & = & \frac{\delta\chi_b [\chi_c(N_I - \bar N) - \bar C]}{\delta + m + \lambda}
\end{aligned}
\end{equation}

where we define 
\be
\gamma = K_c(p) - N_I + \frac{\rho_c \bar C}{\delta}.
\label{eq15}
\ee
Note that these only hold for $\bar B >0$. If $p$ is too low
or $\delta$ too high, the predators will be unable to persist, and
the steady state nutrient and algal densities are 
\begin{equation}
\begin{aligned}
\bar N  &=& \frac{\delta K_c(p)}{ (\chi_c \rho_c -\delta) } \\
\bar C  &=& \chi_c (N_I - \bar N) 
\end{aligned}
\label{eqs9}
\end{equation}

\subsection{Conditions for ESS and ESC}
We now ask under what conditions a particular clone $C_r$, with an associated food value 
$p_r$ might be dominant and non-invasible. Following established usage, a clone $C_r$ 
is said to be an \emph{Evolutionarily Stable Strategy} [ESS] if a population consisting of $C_r$, at steady state, 
cannot be invaded by a rare alternative clone $C_i$ with 
food value $p_i$. The meaning of ``rare'' here is that the growth rate of
a potential ``invader'' $C_i$ is computed on the assumption that the population consists entirely
of the ``resident'' type $C_r$. This quantity is now often called the \textit{invasion exponent}, and
is computed as follows:
\be
\lambda(p_i|p_r) =\frac{1}{C_i}\frac{dC_i}{dt} = \frac{\chi_c \rho_c \bar N}{K_c(p_i) + \bar N}-\frac{G p_i \bar B}{(K_b + \bar C_r p_r)} - \delta 
\label{eqIE}
\ee
where $\bar N$, $\bar B$, $\bar C_r$ are equilibrium conditions set by the resident. 
Note that $\lambda(p_i|p_i) \equiv 0$: pitted against itself, a clone neither increases nor decreases.  

We can characterize ESSs by the first and second derivatives of the invasion
exponent. To be an ESS, an interior trait value $p^*$ (i.e., $p_{min} < p^* < p_{max}$)
must satisfy both the first order condition, $\partial \lambda / \partial p_i = 0 $
and second order condition, $\partial^2 \lambda/\partial p_i^2 < 0$ at $p_i = p_r = p^*$. A 
trait value satisfying the first order condition will be called a \textit{ESS candidate}
(Ellner and Hairston 1994), because the first order condition is necessary
but not sufficient.
For the first order condition we have
\be
\frac{\partial \lambda (p_i | p_r)}{\partial p_i} = \frac{\chi_c \rho_c \bar N \alpha_1 \alpha_2 (1-p_i)^{(\alpha_1 -1)}}{[K_c(p_i) + \bar N]^2} - \frac{G\bar B}{(K_b + \bar C_r p_r)}.
\label{eq17}
\ee
Setting $\bar B^* = \frac{G \bar B}{\chi_c \rho_c}$, we see that $p^*$ is an ESS candidate if  
\be
\Omega_1(p^*)=\frac{\bar N \alpha_1 \alpha_2 (1-p^*)^{(\alpha_1 -1)}}{[K_c(p^*) + \bar N]^2}- \frac{\bar B^*}{(K_b + \bar C p^*)} =0.
\label{eq18}
\ee

Taking the derivative of (\ref{eq17}) with respect to the invader's trait value, 
the second order condition is
\begin{equation}
\frac{\partial^2 \lambda}{\partial p_i^2}  
%& = \frac{\chi_c \rho_c\bar N}{[K_c(p_i) + \bar N]^3}\times\{2\alpha_1^2 \alpha_2^2(1-p_i)^{2(\alpha_1-1)} - \alpha_1 \alpha_2 (\alpha_1 - 1)(1-%p_i)^{(\alpha_1 -2)}[\bar N + K_c(p_i)]\} \\
=\xi_1 \{2\alpha_1\alpha_2(1-p_i)^{\alpha_1} -(\alpha_1 -1)[\bar N + K_c(p_i)]\},
\label{eq19}
\end{equation}
where 
\[
\xi_1 = \frac{\chi_c \rho_c \bar N\alpha_1 \alpha_2 (1-p_i)^{(\alpha_1 - 2)}}{[K_c(p_i) + \bar N]^3} > 0 
\]
provided  $0 \leq p_i < 1.$ Thus, to be an ESS, an interior candidate $p^*$ must also satisfy 
\be
g(p^*) = 2\alpha_1\alpha_2(1-p^*)^{\alpha_1} -(\alpha_1 -1)[\bar N + K_c(p^*)] < 0. 
\label{eq20}
\ee

Where no single ESS exists -- in particular when a candidate fails to satisfy the
second order condition -- it is often possible to identify 
a non-invasible ESC (Evolutionarily Stable Combination) of coexisting genotypes. 
An ESC $\kappa$ of coexisting genotypes is characterized by the invasion exponent for an introduced rare type $p_i$,   
\begin{equation}
 \lambda(p_i|\kappa) =\frac{1}{C_i}\frac{dC_i}{dt} = 
\frac{\chi_C \rho_c \bar N}{K_c(p_i) + \bar N}-\frac{G p_i \bar B}{(K_b + \bar C_\kappa)} - \delta 
\end{equation}
where $\bar N$, $\bar B$, are the steady state nutrient and rotifer densities set by $\kappa$ and
$\bar C_\kappa$ is the ``effective'' algal density at the ESC steady state, 
\be
\bar C_\kappa = \sum\limits_{j \in \kappa} p_j \bar C_j
\ee
If $\lambda(p_i|\kappa) < 0$ for all $p_i \not\in \kappa$ then $\kappa$ is an ESC. A more
refined local classification of evolutionary stability is possible, also based on derivatives
of the invasion exponent (Figure 1 of Levin and Muller-Landau (2000) summarizes the various stability
concepts and their relationships), but the ESS and ESC concepts are sufficient for studying 
our model.

\subsection{Shape parameter $\alpha_1 < 1$}
We observe from (\ref{eq19}) that
\be
\frac{\partial^2 \lambda}{\partial p_i^2} > 0 \mbox{ when } \alpha_1 < 1.
\label{eq23} 
\ee
Therefore no \emph{interior} trait $p$ (i.e., $p_{min} < p < p_{max}$ ) can be an ESS  if $\alpha_1 < 1$.
In addition there cannot be an ESC that includes any interior types (see the Appendix), so any ESC must 
consist of the extreme types $\{p_{min}, p_{max}\}$.
The evolutionary outcome can therefore be determined from the two invasion exponents 
$\lambda(p_{min}|p_{max})$ and $\lambda(p_{max}|p_{min})$ [Table 2]. 

Recall that food value $p$ is scaled
so that $p_{max} \equiv 1$, representing the undefended clone that is favored when predators are absent. 
However, the evolutionary outcome may depend on the value of $p_{min}$, the food value
corresponding to the highest possible level of defense. We consider below a wide range of
possible values for $p_{min}$, i.e. we regard it as being under experimental
control rather than a fixed property of the organism. Although the latter is literally true,
$p_{min}$ can be increased temporarily by using a restricted set of founding genotypes 
as in Yoshida et al. (2003).

\begin{table}[t]
\centerline{\textbf{Table 2. } Predicted evolutionary outcome as a function of the invasion exponents when
$\alpha_1 < 1$.}
\begin{tabular}
{p{120pt}p{120pt}p{120pt}}
& $\lambda(p_{max}|p_{min})>0$ & $\lambda(p_{max}|p_{min})<0$ \\
\hline
\end{tabular}
\begin{tabular}
{|p{120pt}|p{120pt}|p{120pt}|}
%& & \\
$\lambda(p_{min}|p_{max})>0$& ESC $\kappa= \{p_{min},p_{max}\}$ & $p_{min}$ is ESS \\
%& & \\
\hline 
%& & \\
$\lambda(p_{min}|p_{max})<0$ & $p_{max}$ is ESS & Both are (local) ESS \\
%& & \\
\hline 
\end{tabular}
\end{table}

\begin{figure}
\centerline{\includegraphics[width=15.25cm,height=15.25cm]{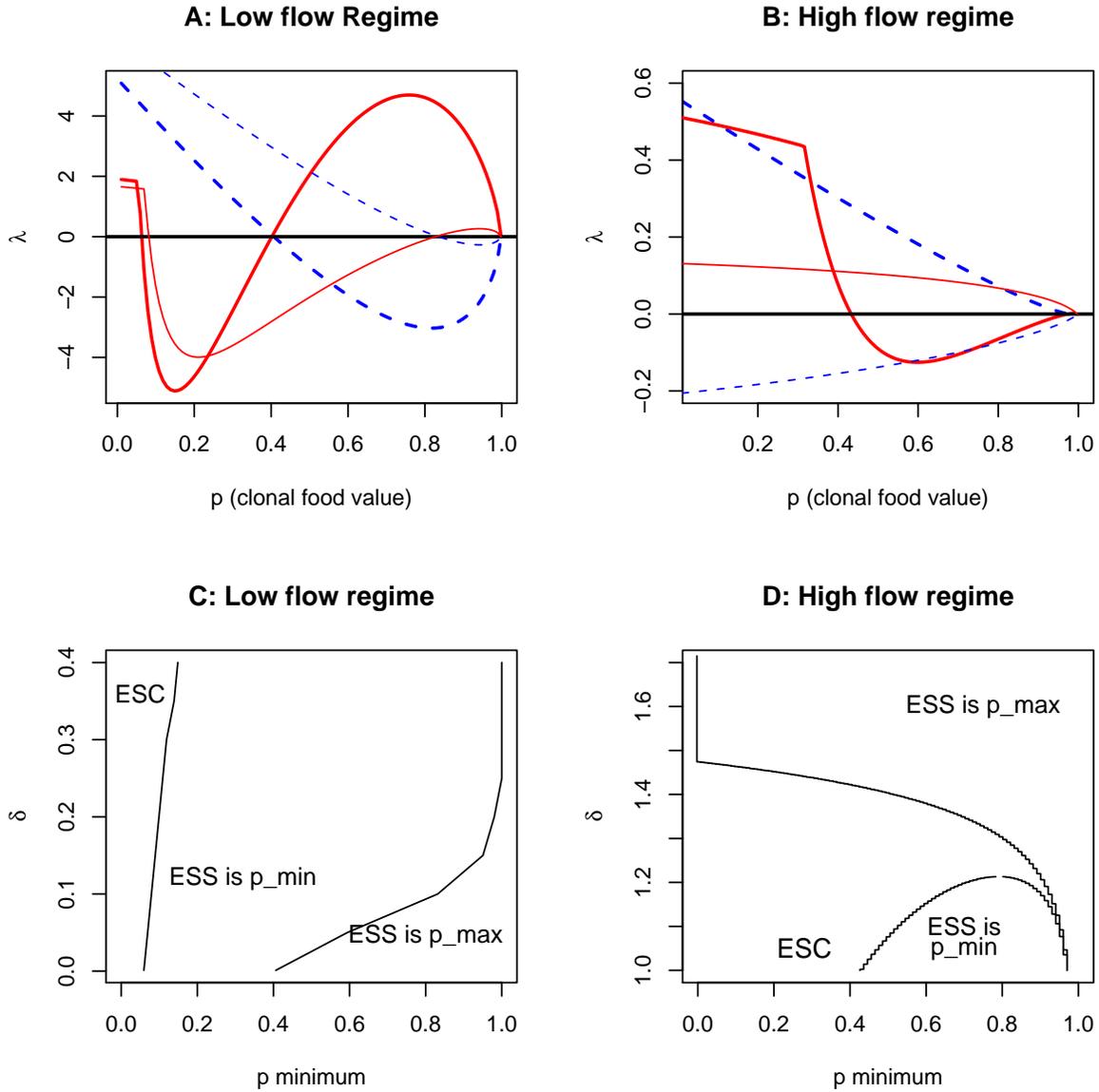}}
\caption{Invasion exponents and evolutionary steady states when $\alpha_1 < 1$. (A) Plots of the invasion exponents
$\lambda(p_{max}|p)$ (solid line) and $\lambda(p|p_{max})$ (dashed line) as a function of $p$ for low flow rates,  
$\delta=0.001$ (bold) and $\delta=0.15$ (thin). (B) As in panel (A) for high values of $\delta$, 
$\delta=1.0$ (bold) and $\delta=1.7$ (thin).  Panels (C) and (D) show the predicted evolutionary steady
states that result from the plots in (A) and (B) using the criteria in Table 2, as a function of
$\delta$ and $p_{min}$. The ``corner'' in the plot of 
$\lambda(p_{max}|p)$ occurs at the minimum food value $p$ that allows persistence of the predators. 
Above this critical value, $\lambda(p_{max}|p)$ is calculated for a system steady state including
$N$, $C$, and $B$; below this critical value only $N$ and $C$ are present in the system at steady
state. } 
\label{Figure3}
\end{figure}

Figure \ref{Figure3} shows plots of the invasion exponents $\lambda(p_{max}|p)$ 
and $\lambda(p|p_{max})$  (panels A and B), and the predicted outcome according 
to the criteria in Table 2 (panels C and D). In the low-flow regime (panel A), 
when $p$ is near 0 both $\lambda(p_{max}|p)$ and $\lambda(p|p_{max})$ are positive. Thus, when $p_{min}$ is in
the range of $p$ values where these inequalities hold, we predict an ESC. 
As $p$ increases, $\lambda(p|p_{max})$ remains positive (dashed line)
but $\lambda(p_{max}|p)$ becomes negative (solid line). 
For $p_{min}$ in this range of $p$ values, we therefore have 
$\lambda(p_{min}|p_{max})>0$ and $\lambda(p_{max}|p_{min})<0$, and the ESS is the minimum existing $p$ 
(that is, the most defended clone). As $p$ increases
further the two exponents reverse in sign, so if $p_{min}$ 
lies to the right of this change-point $p_{max}$ is an ESS. 

The high flow regime is shown in Figure \ref{Figure3}~B.  For $p$ very near 1 the situation
is the same as at low flow: $\lambda(p_{max}|p)>0$ and $\lambda(p|p_{max})<0$, and we again 
have $p_{max}$ as the ESS. 
As $p$ decreases $\lambda(p_{max}|p)$ remains positive for dilution rates at the high end of
the high flow regime, but becomes immediately negative as $p$ decreases from $p$ near 1 at
the low end of the regime, i.e., for $\delta \doteq 1$.
$\lambda(p|p_{max})$ may either remain negative or become positive, depending on the value of $\delta$. 
When a sign-change occurs,  $\lambda(p|p_{max})$ is positive and $\lambda(p_{max}|p)$
negative, and we have $p_{min}$ as an ESS, or both exponents are positive so the predicted outcome is an ESC:
a combination of the two extreme types.
The location of the sign-change decreases with increasing $\delta$, and eventually vanishes. 

The plots shown in Figure 3 are specific to our parameter set, but we show in the Appendix
that the relevant qualitative properties of these curves are robust to substantial variation 
in parameter values.

\begin{figure}
\centerline{\includegraphics[width=15.25cm,height=15.25cm]{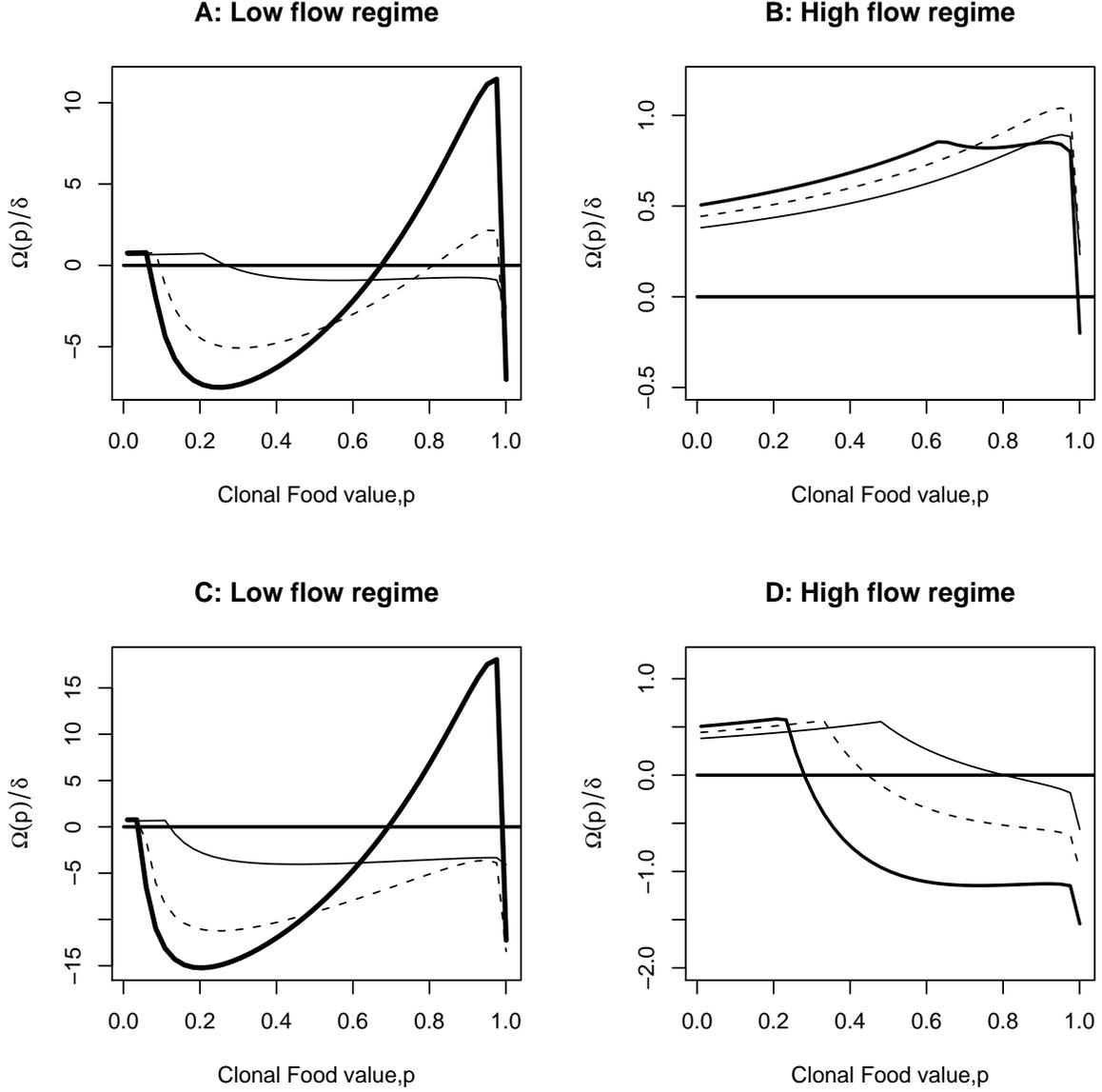}}
\caption{\small First order conditions (\ref{eq18}) for the system when $\alpha_1 >1$ (note that
cost parameter $\alpha_2=9.8$ can be classified as ``high'' in value, so conditions for limiting dilution 
rate $\delta \rightarrow 0$ loop above the zero line). 
A, $\Omega_1(p)$ (\ref{eq18}), scaled by dilution rate, $\delta$ is 
shown for the low dilution regime ($\delta=0.1-0.5$) and conversion efficiency $\chi_b=4000$ (thin line). The limiting form 
for $\delta \rightarrow 0$ is shown in bold line; representative curves for $\delta=0.1$ and $\delta=0.5$ are shown in thin dashed and
solid line, respectively. B, $\Omega_1(p)$ for high dilution rates $\delta = 1.0-1.5$, $\chi_b=4000$; 
curve for $\delta=1.0$ is shown in bold; representative curves for $\delta = 1.25$ and $\delta=1.5$ are shown in thin dashed
and solid line, respectively. C, As in panel A for $\delta=0.1-0.5$, $\chi_b=6000$. Again, 
the limiting form for $\delta \rightarrow 0$ is shown in bold. D, As in panel B for $\delta=1.0-1.5$ and  $\chi_b=6000$.
Plots for $\delta=1.0$ shown in bold.  Note that in the high dilution rate regime, the ESS candidate values move from left to right as  dilution rate increases.} 
\label{Figure4}
\end{figure}

\subsection{Shape parameter $\alpha_1 > 1$: analysis}
The situation is more complicated if $\alpha_1 > 1$, because an internal
ESS is then possible. We therefore need to analyze (in this Subsection) properties of the 
first and second order conditions for a local ESS. The implications for evolutionary
steady states when $\alpha_1 > 1$ are presented in the next Subsection.   

\subsubsection{\it First order conditions}
 
We return to the first order expression,  $\Omega_1(p)$ (\ref{eq18}), and 
search for ESS candidate types by exploring where (\ref{eq18})
holds as a function both of clonal food value $p^*$ and 
dilution rate $\delta$, 
given $0 \leq p^* < 1$ and $ \alpha_1 > 1 $. 
Example calculations of $\Omega_1(p)$ as a function of trait value $p$, 
dilution rate $\delta$, and rotifer conversion efficiency $\chi_b$
are shown  in Figure \ref{Figure4}. 
Each zero point in a curve corresponds to a ESS candidate associated with a given dilution rate,
$\delta$.   

For very low dilution rates,  $\Omega_1(p)$ converges to a limiting form shown in bold (Figure \ref{Figure4}~A,C; bold line), for which, in this case, there are three ESS candidates. Depending on the value of the cost parameter, $\alpha_2$, 
the limiting form may either remain below the zero axis (low $\alpha_2$ values, one ESS candidate), or loop above it (high $\alpha_2$ values, three candidates) as  in Figure \ref{Figure4}~A. As dilution rate increases, the curve dips entirely 
below the $p$-axis (Figure \ref{Figure4}~A,C; thin dashed and solid line), and the two higher--$p$ candidates move together, eventually colliding and vanishing.  The trait value(s) of ESS candidates are little affected by variations in rotifer conversion efficiency $\chi_b$ in the low dilution regime (Figure \ref{Figure4}~A,C).  Note that both $\delta$ and $\chi_b$ enter the first order 
condition, $\Omega_1(p)$, through the steady state values $\bar N$, $\bar B$, and $\bar C^*$ (\ref{eqs8}).

In the high dilution rate regime, there is a single ESS candidate which increases
in $p$--value as dilution rate increases (Figure \ref{Figure4}~B,D). Lower estimates of rotifer conversion efficiency result in higher $p$-value ESS candidates. For example, at $\chi_b=4000$, an ESS candidate exists at $p \approx 1$ for $\delta=1.0$ (Figure \ref{Figure4}~B, heavy line).  However, at $\chi_b=6000$, the candidate for the dilution rate $\delta=1.0$ sits at $p \approx 0.3$ (Figure \ref{Figure4}~D, heavy line).  Higher conversion efficiencies result in greater numbers of rotifers, increased selection for defense, and thus a lower $p$-value ESS candidate.  

Most noticeably at higher dilution rates, varying $\alpha_1$ also affects the position of the ESS candidat. If $\alpha_1$ is increased from some initial
value $\alpha_1 > 1$, candidate $p$ values shift to the left.  
This is sensible when one considers the effect of an increase or decrease in $\alpha_1$ on the trade-off curve.  Increasing $\alpha_1$ makes the curve more deeply concave upwards, while decreasing $\alpha_1$ flattens it. As the curve becomes more deeply concave upwards,
one might expect surviving clones to migrate in $p$-value towards the middle of the curve, and as it flattens, to move to the extremes -- thus, for high $p$-values, to shift towards the high-$p$ endpoint.

\subsubsection{\it Second order conditions}

Second order conditions (\ref{eq20}) are strongly affected by variation in several parameters, most notably shape parameter $\alpha_1$, rotifer conversion efficiency $\chi_b$, and dilution rate $\delta$. 
In the Appendix we demonstrate that $g$ is an decreasing function of both $\alpha_1$ and $\chi_b$; a decreasing function of $\delta$ in the low dilution range, but an increasing function of $\delta$ in the high dilution range.  Consequently, 
increases in either $\alpha_1$  or $\chi_b$ shift the second order curve to the left (at a given dilution
rate, $\delta$), so that lower $p$ value
candidates satisfy the second order condition.  Conversely, decreasing either $\alpha_1$ or $\chi_b$ 
will shift the curve to the right, so that only high $p$ value candidates, can satisfy the second order condition.

The behavior of $g$ with increasing $\delta$ implies that ESS candidates must shift to the right as a function of $\delta$ in order to satisfy the second order condition (\ref{eq20}). All else being held constant, for $\alpha_1 >1$, only relatively high--$p$, high--$\delta$ candidates thus satisfy both the first order (\ref{eq18}) and second order conditions. An internal ESS may exist at high $\delta$ equilibria, depending on conversion efficiency $\chi_b$, but not at low $\delta$ equilibria.

\begin{figure}
\centerline{\includegraphics[width=15.25cm,height=7.62cm]{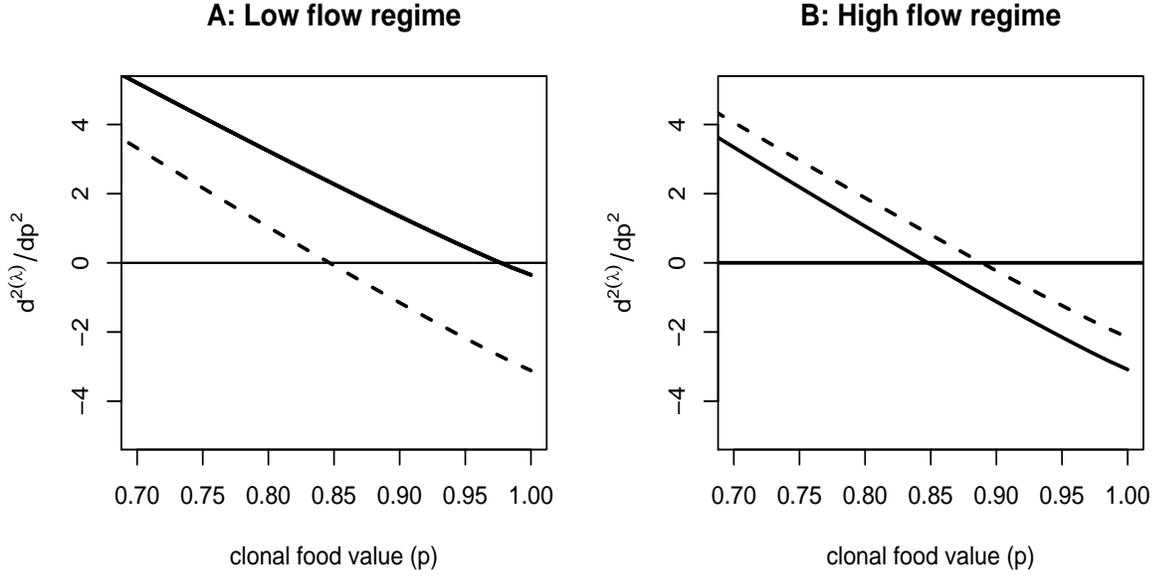}}
\caption{\small Second order conditions (\ref{eq20}) for the system when $\alpha_1 >1$. Curves represent second order conditions as a function of $p^*$ for  equilibrium dilution rates $\delta = (0.1-0.5;1.0-1.5)$. A: second order conditions for low dilution rate regime $\delta=0.1-0.5$. Dashed lines show curves for $\delta =0.1,0.5$.  Heavy bold line shows the limiting form (\ref{eq20}) assumes at very low $\delta$ (i.e., $\delta < O(10^{-3})$).  B: second order conditions for the high dilution rate regime $\delta=1.0-1.5$. In bold line is shown the curve for $\delta=1.0$; dashed line shows curve for $\delta =1.5$ }
\label{Figure5}
\end{figure}

Plotting (\ref{eq20}) as a function of p and $\delta$,  we see that low--$\delta$, low--$p$ candidates invariably fail the second order condition,
because $g(p^*)$ remains positive for these candidates (Figure \ref{Figure5}). For very low (limiting) $\delta$, there may be high--$p$ candidates which satisfy the second order condition; cf. Figure \ref{Figure4}, panels A, C, bold line. The presence or absence of these candidates depends primarily on cost parameter $\alpha_2$, as discussed in the next section.   However, as $\delta$ increases, the ESS candidate shifts to higher $p$--values, and for these values of $p^*$, $g(p^*)$ becomes negative.  Thus while an internal ESS may exist at high dilution equilibria, there is no internal ESS at low dilution rates: a low dilution rate ESS, if it exists, must be an endpoint value.

\subsection{Shape parameter $\alpha_1 > 1$: Results}
\begin{figure}
\centerline{\includegraphics[width=15.25cm,height=15.25cm]{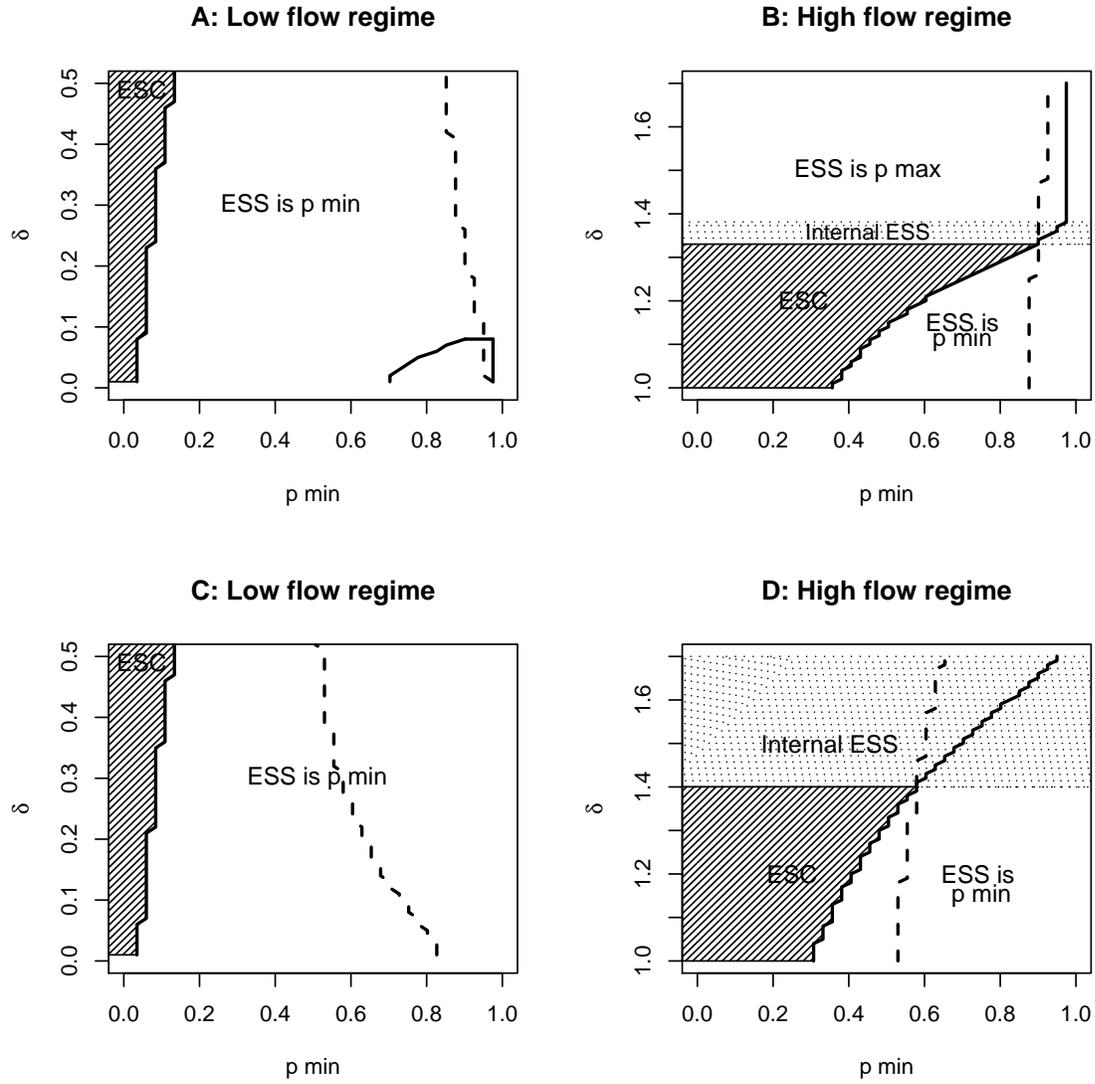}}
\caption{Evolutionary steady states when $\alpha_1 >1$. ESS candidates are shown in solid line; 
to the {\em right} of the dashed line the second order condition 
$\frac{\partial^2 \lambda}{\partial p_i^2} < 0$ is satisfied, so a candidate is an ESS if it lies to the right of the
dashed line. If $p_{min}$ lies to the left of the ESS candidate line, the evolutionarily stable state is either an
ESS (if the candidate satisfies the second order condition) or an ESC (if the candidate does not satisfy this condition).
Parameters for system shown in panels A-B: $\alpha_1 = 1.08$, $\alpha_2= 9.8$, $\chi_b=5420$; panels C-D: $\alpha_1=1.19$, $\alpha_2=3.9$, $\chi_b=6055$.}
\label{Figure6}
\end{figure}

We now use the results above to identify the form of the evolutionary steady states for  $\alpha_1 > 1$. 
Figure (\ref{Figure6}) summarizes the results.   
Panels A through D show first and second order conditions for this system together on one plot: 
solid lines indicate ESS candidates, which we'll call $p^*$,  
and the dashed line is a plot of the second order condition  
$g(p^*)$ (\ref{eq20}). To the left of the dashed line, 
ESS candidates fail the second order condition; to the right of this line, candidates satisfy this condition.  
Note that first order conditions for the parameters in \ref{Figure6}~A, B are found in
\ref{Figure4}~C, D.

For a low flow regime with high cost parameter value $\alpha_2$ and for $p$ small,
there are ESS candidates, $p^*$ (solid line, panel A), but they fail the second order condition. 
Then, for $p_{min} < p^*$, 
there is thus an ESC of the two extreme types in proportions such that the ESC has
an average $p$ value approximating the ESS candidate value, $p^*$.  For $p_{min} > p^*$, 
the first order condition becomes negative and all clones are invasible by $p_{min}$, which forms an endpoint ESS.  
For the high cost-value ($\alpha_2\doteq 10$) example shown in
panel A, there is an additional set of high $p$ ESS candidates at very low dilution rate. 
However, simulations indicate that these candidates form only a very local ESS and may be
invaded by lower $p$ clones. Note that if the cost parameter is lower 
(e.g., $\alpha_2= 4$, panel C),
these high $p$, low $\delta$ ESS candidates do not occur, and only the ESC and
low-$p$ endpoint ESS are observed.

In the high-flow regime, ESS candidates $p^*$ generally increase in $p$-value with
increasing dilution rate (panels B and D, solid line). For this parameter set, however, 
all ESS candidates for $\delta < 1.35$ fail the second order condition.
Thus, for $p_{min} < p^*$, there is again an ESC of $p_{min}$ and $p_{max}$, in proportions such that the
average $p$ value approximates the candidate value $p^*$ (panel B).  For $p_{min} > p^*$, the first 
order condition becomes negative, so higher $p$ clones are invasible by $p_{min}> p*$, and $p_{min}$ 
is again an endpoint ESS. However, where ESS candidates do not fail the second order condition, in 
this case for $\delta > 1.35$, an internal ESS exists (above dotted line, panel B). 
Recall that as $\alpha_1$ is increased, we have shown that the second order curve $g(p^*)$ shifts 
to the left.  In this example this causes the intersection between the first and second order curves 
(dotted line, panel D), 
to occur at lower and lower dilution rates. Thus more and more of the high dilution rate 
regime has an internal ESS. 

Recalling that shape parameter $\alpha_1$ and assimilation efficiency $\chi_b$ affect first and second 
order conditions at high dilution rates, we turn to a second example (panels C,D).  Here both $\alpha_1$ and $\chi_b$ have been increased.   
Increasing $\alpha_1$ shifts ESS candidate values to the left, as does increasing $\chi_b$. 
However, this example also features a lower cost parameter ($\alpha_2 \sim 4$), which {\it steepens} 
the first order line such that all ESS candidate types fail the second order condition for these parameters. 
Thus as above, for $p_{min} < p^*$, there is an ESC centered about
the ESS candidate value, and for $p_{min} > p^*$, $p_{min}$ forms an endpoint ESS.  Since ESS candidates fail 
the second order condition for all dilution rates, an internal ESS never exists.  The results are otherwise 
the same as in panel B.

\section{Simulations}
Since results from the type of ESS analysis we have just completed may only hold locally, we 
tested our analytical conclusions by performing simulations of the system (\ref{eq1})-(\ref{eq6}). The  simulations were written
in the \textbf{R} language (version 1.7) and run on a Windows 2000 platform, using the \textbf{odesolve} package.  
Tests were performed for a representative sample of parameter sets with 
$\alpha_1 < 1$ and $\alpha_1 > 1$ and in a range of dilution rates, as follows.

First, summary diagrams, such as those shown in Figure \ref{Figure3}~C,D (for $\alpha_1 < 1$) and
Figure \ref{Figure6}~A-D (for $\alpha_1 > 1$), were created to show the predicted evolutionary equilibria for each
parameter set. Then, each diagram was ``gridded" coarsely in ($p_{min}$, $\delta$) and 
simulations of the system in equations (\ref{eq1})-(\ref{eq6}) were run to verify the analysis 
at each grid point. Some additional ``spot-check'' simulations were used to look
more closely at predicted transition points, e.g. from ESC to ESS. In each case, the system included an initial 
set of 40 clones, distributed evenly in $p$ value along the tradeoff curve determined by the parameter set in question. 
After running a given simulation, we examined time-series trajectories for $C$, total {\it Chlorella} cell counts;
average palatability $p$; normalized clonal frequency $c_i$ as a function of time; and mean clonal frequency as a function of
$p$.  Each run verified that (1) the system was in equilibrium ($p_{min}$ is a bifurcation parameter, thus equilibrium must be repeatedly
confirmed as $p_{min}$ varies); and (2) the clone or clones identified by analysis were indeed dominant. 
In only one case, noted above, did we find that an ESS identified by our analysis was 
only local in nature, and thus invasible by clones of much lower $p$ value.

\section{Evolution and response to changing conditions}

\begin{figure}
\centerline{\includegraphics[width=15.25cm,height=15.25cm]{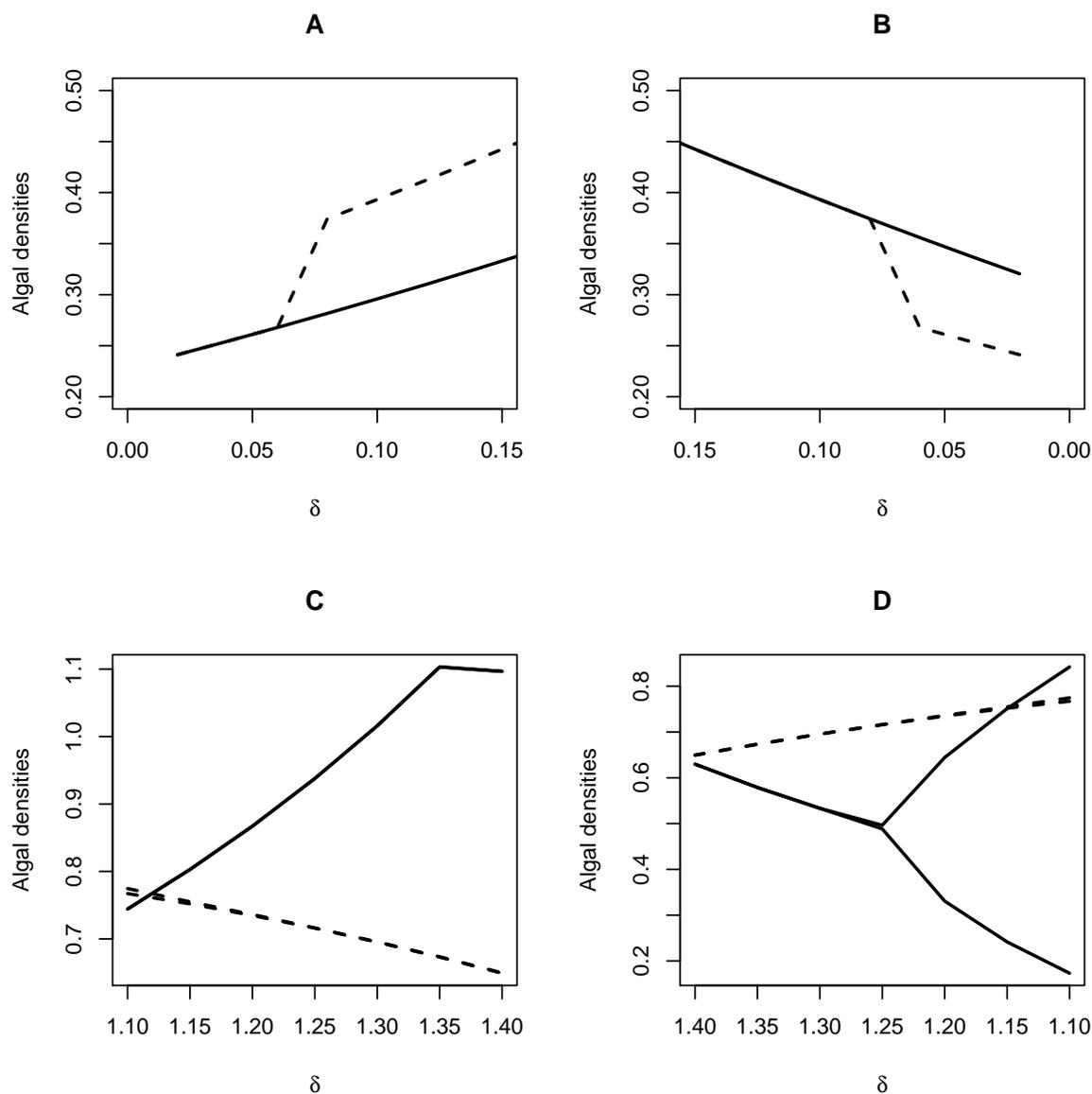}}
\caption{Effects of change in dilution rate on steady-state algal density 
in an evolutionarily dynamic system versus a
static (non-evolving) system, for $\alpha_1 < 1$ (panels A, B) and $\alpha_1 > 1$ (panels C, D). 
Panels A, B: $p_{min} \doteq 0.75$; solid line shows static response, and dashed line shows dynamic response. 
Parameters and evolutionary dynamics are as in Figure \ref{Figure3}~C.
A, Initial dilution rate $\delta=0.02$, increased to $\delta=0.15$. 
B, Initial dilution rate $\delta=0.15$, decreased to $\delta=0.02$.  
Panels C, D: parameters and evolutionary dynamics are as in Figure \ref{Figure6}~D.
The dynamic system (dashed line) includes 40 clones; the static system (solid line) is a single-clone
whose parameters reflect the aggregate properties of the ESC. 
C, Initial dilution rate $\delta=1.1$, increased to $\delta=1.4$.
D, Initial dilution rate $\delta=1.4$, decreased to $\delta=1.1$.
}
\label{Figure7}
\end{figure}

We can now return to the question posed in the Introduction: how important is an understanding of 
evolutionary dynamics for making quantitative predictions of how the system will respond to ``management'' 
(an externally imposed alteration in conditions)? If, as the preceding analysis and examples suggest, 
a change in conditions results in a change in community composition, then predictions of how such a community might 
respond to management must not omit  the effects of evolution. 

Figure 7 shows predicted responses of the system with and without evolution to changes in the dilution rate,
which affects both nutrient supply and mortality rates. 
In panels A,B we assume  $\alpha_1 < 1$, and
evolutionary dynamics are as summarized in Figure \ref{Figure3}~C. In both panels, the dashed line shows the effects on {\it Chlorella} density assuming the system responds to external change by evolving, while solid line shows the predicted response if effects of evolution are ignored.
In panel A we set $p_{min}=0.75$, then slowly increase the dilution rate 
from $\delta=0.02$ to $\delta=0.15$,
thus crossing from ``ESS is $p_{max}$'' to ``ESS is $p_{min}$'' on Figure \ref{Figure3}~C.
Both systems respond identically initially, but at the ``transition'' dilution rate ($\delta \doteq 0.8$ for this choice of parameters), population densities in the evolutionarily dynamic 
treatment suddenly increase to a higher steady-state (dashed line), while those in the static treatment follow the 
original (solid line) steady-state.  In both systems the initial dominant was $p_{max}$, but only the evolutionarily dynamic system shifted to the more competitively fit dominant type, $p_{min}$.

In panel B we run the converse experiment: we set $p_{min}= 0.75$, but start at dilution rate
$\delta=0.15$.  We then decrease the dilution rate slowly back down to $\delta=0.02$, thus passing back from ``ESS is $p_{min}$''(Figure \ref{Figure3}~C) to ``ESS is $p_{max}$''. Both systems begin with 
$p_{min}$ as dominant and their response is identical, but as dilution rate passes through the transition at $\delta \doteq 0.08$, the evolutionarily dynamic system switches to $p_{max}$ as dominant, resulting in lower algal densities (dashed line).  The evolutionarily static system continues to be dominated by $p_{min}$ (solid line).
So up to a point, the system's response to changing conditions can be predicted without regard to the underlying 
evolutionary dynamics, but then a rapid evolutionary response to a small change in conditions causes a discontinuous
response that would not be predicted when evolutionary dynamics are ignored.  

With $\alpha_1 >1$ (Figure \ref{Figure7}C,D) there is an immediate divergence between the actual system response, 
and predictions that ignore the effect of evolution. 
Again, in both panels, the solid line shows the response of an 
evolutionarily static system, and in dashed line is the response of the evolutionarily dynamic, multiple 
clone system. In panel C we choose an initial dilution rate of $\delta=1.10$, and slowly increase the dilution rate to $\delta=1.4$. 
At the initial dilution rate, the multiple clone system has an ESC
comprised of the extremes, $p_{min}$ and $p_{max}$, in proportions such that their 
average $p$ value approximates the ESS candidate at that dilution rate. Algal densities (dashed line) 
slowly fall as dilution rate increases, due to an increase in average $p$ value of the ESC with 
increasing dilution rate (higher $\delta$ reduces the rotifer density, and thus reduces the 
selection for unpalatability). The evolutionarily static system has a single clone,
whose parameters reflect the aggregate properties of the ESC (for example, its value of $K_c$ was set 
to the nutrient concentration at which the total uptake rate of the ESC clone mix reached half its maximum value). 
Here, higher dilution does not evoke the evolutionary response of increased palatability in the prey, so 
rotifers begin to drop out of the system almost immediately, with a consequent steady increase in algal 
densities (solid line). By $\delta = 1.35$, the rotifers are extinct, and algal densities stabilize.

We then run our imaginary experiment the other way (Figure \ref{Figure7}~D), selecting 
an initial dilution rate of $\delta=1.4$, and  slowly decreasing the dilution rate back 
down to $\delta=1.1$.  The dynamic system responds to a decrease in dilution rate with a 
slow increase in algal densities: this is because the average palatability of clones 
comprising the ESC is decreasing with $\delta$, with the expected effect on rotifer abundances.  
The static system (a single clone mimicking the aggregate properties of the ESC at $\delta=1.4$) 
maintains constant palatability as conditions change, and again its response is very different 
from the outcome when evolution occurs (note that the system begins to cycle as we pass
back through a bifurcation point at $\delta \doteq 1.25$).

\begin{figure}
\centerline{\includegraphics[width=15.25cm,height=15.25cm]{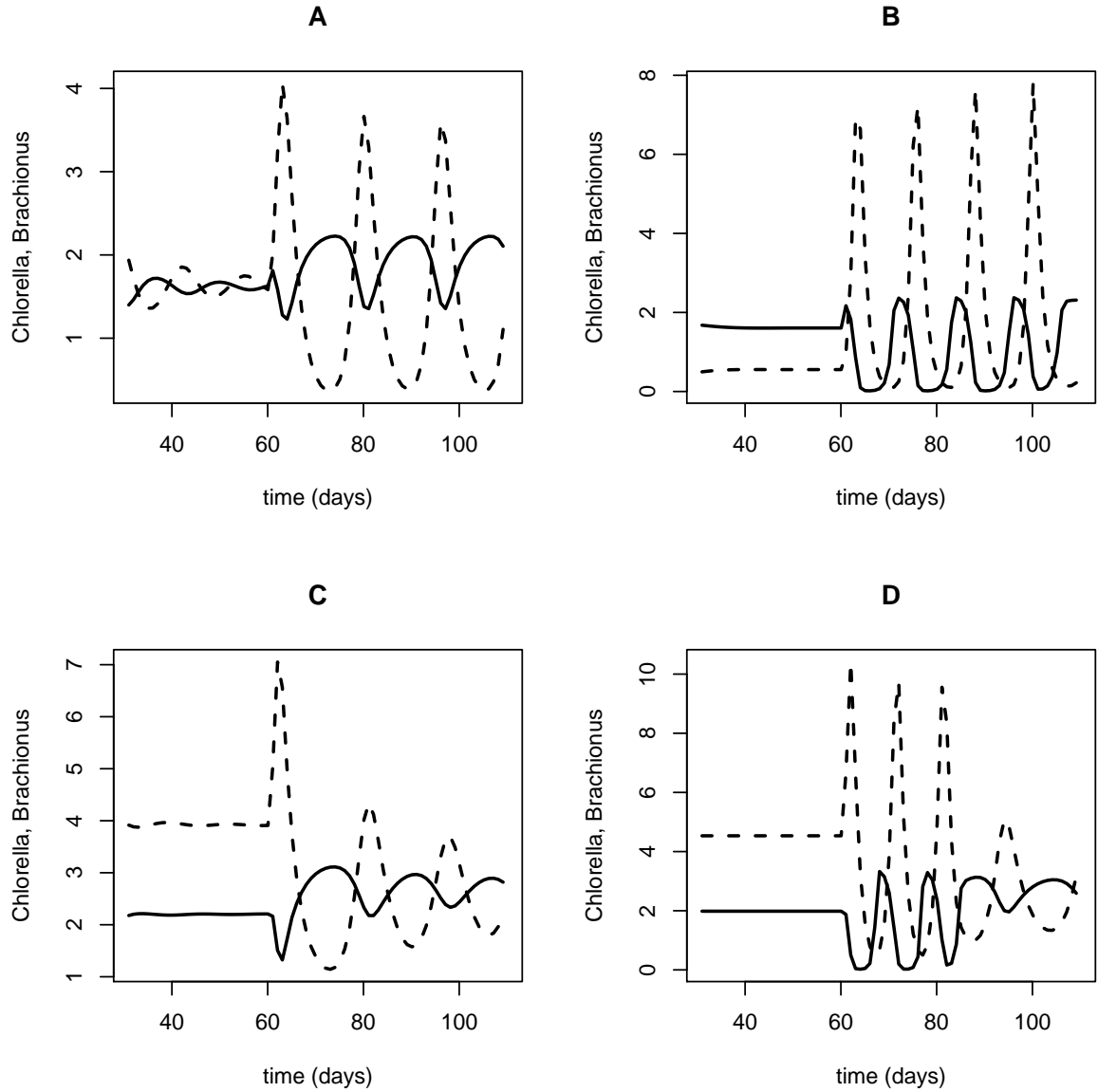}}
\caption{Effects of change in dilution rate on an evolutionarily dynamic system.
Panels A, B: for $\alpha_1 < 1$. Evolutionary dynamics are as in Figure \ref{Figure3}~D; parameters are as follows:
$\alpha_1 = 0.765$, $\alpha_2 = 8.23$, $p_{min}=0.07$, $\chi_b=6400$. 
Panel A: Initial dilution rate $\delta=1.15$ stepped down to $\delta=0.7$ at $t=60$;
B, Initial dilution rate $\delta=1.65$ stepped down to $\delta=0.7$ at $t=60$.
Panels C, D: for $\alpha_1 > 1$. Evolutionary dynamics are as in Figure \ref{Figure6}~B;
parameters $\alpha_1 = 1.087$, $\alpha_2 = 9.79$, $p_{min}=0.12$, $\chi_b=5419$.
Rotifers are shown in dashed line and {\it Chlorella} in bold line.
Panel C, Initial dilution rate $\delta=1.2$ stepped down to $\delta=0.7$ at $t=60$;
D, Initial dilution rate $\delta=1.45$ stepped down to $\delta=0.7$ at $t=60$.}
\label{Figure8}
\end{figure}

We now propose a set of experimental checks for our analysis that are robust to 
parameter uncertainties, and in particular to the shape of the tradeoff curve. 
A general prediction for any plausible parameter values
is that the evolutionary steady-state at very high dilution rates is a single
clone (either $p_max$ or an ESS near $p_max$), which is replaced by an ESC
as the dilution rate decreases. As dilution rate decreases further, 
a Hopf bifurcation occurs and the attractor is then a limit
cycle. The nature of the cycles (their period and the phase lag between predator
and prey) depends on which prey clones are present (Yoshida et al. 2003). 

We therefore predict that if the dilution rate is quickly dropped into the
limit-cycle regime, the nature of the cycles will (at least initially) depend on
the prior dilution rate and evolutionary steady state. Figure \ref{Fig8} shows 
this scenario for both $\alpha_1 < 1$ (panels A, B)
and $\alpha_1 > 1$ (panels C, D). In panel A ($\alpha_1 < 1$),
we start with an initial dilution rate of $\delta=1.15$. The predicted outcome
is an ESC of the extreme types, $p_{min}$ and $p_{max}$ (Figure \ref{Fig3}~D).  
After equilibrium is established, we abruptly step down the dilution rate 
to $\delta=0.70$. Given the presence of two very different clones, we predict longer cycles with
algal and rotifer peaks exactly out of phase (Yoshida et al. 2003).
In panel B we start with dilution rate $\delta=1.65$, at which we expect an
ESS of $p_{max}$ (Figure \ref{Fig3}~D). When the dilution rate is stepped down to $\delta=0.70$, 
short cycles with classical predator-prey phase relations result, as one would expect from a single-clone
system (Yoshida et al. 2003). Panels C and D show a corresponding step-down experiment 
for $\alpha_1 > 1$. In both cases, the nature of the cycles that occur immediately after the step-down
reveals the genetic diversity that was present before the step-down, and thus lets us
test our predictions about the ``hidden'' prey diversity at steady-state. Note, however, that
prey evolution continues to occur after the step-down (panel D), so only the transient
behavior after the step-down is revealing of prior evolutionary history.

\section{Discussion}
Our analysis of evolutionary dynamics suggests that small, gradual changes in external conditions can 
precipitate dramatic shifts in community composition or species' densities. This can occur for 
either qualitative type of tradeoff curve (Figure \ref{Figure1}), through two different mechanisms. 

For $\alpha_1 < 1$, extreme prey types are favored and an internal ESS cannot exist. Instead, the prey 
population is dominated either by an endpoint ESS (the minimum or maximum possible food values), or an ESC 
consisting of the two extreme types. Under a small change in conditions, the evolutionary steady state can 
jump from one endpoint ESS to another, resulting in a discontinuous change in species abundances 
(Figure \ref{Figure7}AB). 

For $\alpha_1 > 1$, an internal ESS may be present at high dilution rates, 
depending on the shape and cost parameters $\alpha_1$ and $\alpha_2$, and rotifer conversion efficiency $\chi_b$. 
Where an internal ESS does not exist, the population is dominated either by an endpoint ESS composed of the 
minimum food value clone present in the system, $p_{min}$, or an ESC of coexisting types. Simulations 
indicate that the ESC consists of the two extreme types $p_{min}, p_{max}$ in proportions such that the
average food value of the population approximates that of the ESS candidate (which satisfies the first
order condition but not the second order condition for evolutionary stability). Because the ESC
consists of two very different types, the balance between them is strongly affected by relatively
small changes in conditions. As a result, the response of the system (including evolutionary
changes in the prey) is very different, not only in magnitude but in direction, from what would be
predicted if evolutionary changes in the prey are ignored. 

Experiments are in progress to determine the shape of the tradeoff curve in our system. Circumstantial evidence 
supports a tradeoff curve where extreme types are favored ($\alpha_1 < 1$):
\vspace{-0.2in} 
\begin{itemize}
\item The best fits of the model to qualitative properties of the experimental data (in particular the 
transition points between stability and cycles as a function of dilution rate, and properties of 
the cycles) are obtained with $\alpha_1 < 1$. 
\item The variability in cycle period (roughly 20-40d) observed in experiments can occur in the model when 
$\alpha_1 < 1$ (depending on which clones are present), but not for $\alpha_1 > 1$ (Yoshida et al. 2003). 
\end{itemize}
\vspace{-0.2in} 
This raises the possibility that a rapid jump from one extreme prey type to another, as predicted by the model,
may be experimentally observable in our system. 

Our analysis complements recent work by Abrams and Vos (in press) on a general model for 
a resource-consumer-predator food chain, with adaptive change  
in a consumer trait affecting food consumption and mortality rates. 
Their goal was to predict how perturbations at one trophic level - such 
as an increase in predator mortality - would propagate through the chain and alter 
other species' abundances, in the presence of consumer adaptation. The predictions from purely 
ecological models that ignore consumer adaptation have sometimes been supported, but are also 
contradicted by numerous experimental studies (Abrams and Vos in press). Abrams and Vos 
(in press) showed that consumer adaptation broadens the range of theoretically possible responses, and so 
might explain some cases where purely ecological models failed. For example, without adaptation an incremental 
increase in consumer mortality in their model always entails an increase in resource abundance and a 
decrease in predator abundance; with consumer adaptation the resource abundance may decrease and predator
density may decrease. 

Abrams and Vos (in press) consider a spectrum of models with different assumptions about
the composition of the consumer trophic level and the nature of response -- e.g., assuming
a single phenotypically homogenous consumer species responding behaviorally, or a pair of
consumer species changing in relative abundance -- and found that predictions about total
consumer abundance were robust across a range of models. In our analysis the
composition of the consumer trophic level is not assumed, but is one of the predicted
outcomes from the clonal selection dynamics. As a result, we can predict how
prey evolution can lead to qualitative changes in the genetic composition of the prey population, 
and how these qualitative changes can lead to abrupt population responses to
gradual changes in conditions, as well as the gradual changes predicted by the Abrams-Vos 
models. 

Working with well-characterized model ecological systems (e.g., Mueller and Joshi 2000, 
Cushing et al. 2002) makes it feasible to study processes that would be far less tractable 
in the field, and less amenable to rigorous testing of theoretical predictions. Our results  
lay the groundwork for rigorous tests of the longstanding  but still contentious hypothesis 
that population management must be ``evolutionarily enlightened'' (Ashley et al. 2003), rather
than continuing to take an exclusively ecological perspective (Stockwell et al. 2003). 

\appendix
\section{Appendix}
\subsection{$\alpha_1<1$}
To see that we cannot have an ESC including any interior types, consider a coexisting set $\kappa$ of two or more genotypes,
which has an interior member $l$ and some other member $j$.
By definition we have $\lambda(p_l|\kappa) = \lambda(p_j|\kappa)=0$, and as a result of the positive
second derivative (\ref{eq23}) we conclude that $\kappa$ must be invasible by some other interior
genotype near $p_l$. Thus, $\kappa$ as defined cannot be an ESC.  

We now analyze the qualitative properties of the invasion exponent 
functions for $\alpha_1<1$. 
Consider first $\lambda(p_{max}|p)$ for low flow-rate $\delta$ (Figure \ref{Figure3}A). 
For a resident type with $p$ near 0 we have $\bar B=0$ (the predators cannot persist) 
so defense against predation has no value and any clone with higher $p$ can invade, 
implying that $\lambda(p_{max}|p)>0$. As the food value $p$ of the
resident type increases, the predators can then persist (this bifurcation accounts for the corner in the plot 
of $\lambda(p_{max}|p)$). We can approximate $\lambda(p_{max}|p)>0$ for $\delta$ small, 
by inserting the Taylor series expansions of the steady-state values ($\bar N$, $\bar B$, $\bar C$) 
about $\delta = 0$ into (\ref{eqIE}). This gives 
\[
\delta^{-1}\lambda(p_{max}|p) = \frac{1}{p} - \theta_1 + O(p) + O(\delta)
\]
where
$\theta_1 = N_I\chi_c[\chi_b G - (m + \lambda)]/[K_b (m + \lambda)].$ 
The estimated value of $\theta_1$ from our experimental data, depending on the value of $\chi_b$, is
in the range $16.3 - 26.8$, causing $\lambda(p_{max}|p)$
to become negative roughly at $p= \theta_1^{-1} \doteq 0.04 - 0.06.$ Thus, the qualitative behavior
of $\lambda(p_{max}|p)$ at small $p$ is robust to any parameter changes such that  $\theta_1 \gg 1$
continues to hold. 

Similarly, for $\delta^{-1}\lambda(p|p_{max})$ near $p=0$ we obtain
\[
\delta^{-1}\lambda(p|p_{max}) = \theta_1 \frac{K_c(p_{max})}{K_c(p)} - (\theta_1-1) p -1 + O(\delta).
\]
Thus $\lambda(p|p_{max})>0$ for $p$ small if $\frac{K_c(p_{max})}{K_c(0)} > 1/\theta_1$,
which is true for our estimated parameters (the LHS is at least $\approx 0.25$, the RHS no more
than $\approx 0.06$). However, as $p$ increases the $\theta_2$ term dominates, and $\lambda(p|p_{max})$ becomes negative,
as seen in Figure \ref{Figure3}A. As $\delta \to 0$ both scaled invasion exponents 
approach limiting shapes similar to the curve shown in bold (Figure \ref{Figure3}A).

For low and high flow regimes, as $p \to 1$ the slopes of both invasion exponents become
infinite (Figure \ref{Figure3}AB). With some algebra we can show that this feature depends only
on the value of $\alpha_1$.  In the limit as $p \to p_{max}=1$, from (\ref{eq17}) we have
\[
\frac{\partial\lambda(p_{max}|p)}{\partial p} \sim + \textrm{Constant}\times \frac{\partial K_c}{\partial p}
\]
and
\[
\frac{\partial\lambda(p|p_{max})}{\partial p} \sim  - \textrm{Constant}\times \frac{\partial K_c}{\partial p}.
\]
Since $\frac{\partial K_c}{\partial p} = - \alpha_1\alpha_2(1-p)^{\alpha_1 -1}$ and  
$\alpha_1 < 1$, then 
$\frac{\partial\lambda(p_{max}|p)}{\partial p} \to -\infty$
and
$\frac{\partial\lambda(p|p_{max})}{\partial p} \to +\infty$
as $p \to p_{max}.$ 

\subsection{$\alpha_1>1$}
Here we derive the properties of the function $g(p)$ defining the second-order condition, that
were used in the case $\alpha_1>1.$ As usual these properties depend on parameter values and
our goal is to identify which parameters or parameter combinations control the relevant
properties. 

\large{(a) $g(p^*)$ is a decreasing function of $\alpha_1$ for all $\alpha_1>0$}\\
\normalsize
Substituting the steady state expressions (\ref{eqs8}) into (\ref{eq20}), 
we rewrite $g$ as a function of $\alpha_1$. Taking a partial derivative of this expression with respect to $\alpha_1$ yields
\[
\frac{\partial g(p^*)}{\partial \alpha_1} = \frac{K_c(p)'}{2}\left[ (3\alpha_1 - 1)- (\alpha_1 - 1)\frac{\gamma + 2N_I}{\sqrt{\gamma^2 + 4N_I K_c(p^*)}}\right] - \left[ \bar N + K_{c_{min}} - \alpha_2(1-p)^{\alpha_1} \right]
\]
where $\gamma$ is defined in equation (\ref{eq15}), and 
$K_c'(p)= \frac{\partial K_c(p)}{\partial \alpha_1} = \alpha_2 ln(1-p)(1-p)^{\alpha_1}$. 
 
Note that $K_c'(p)$ is negative for any $0 < p < 1$. The expression in the first of the two square brackets is
always positive and remains bounded  between $\approx 1.4 - 2$ as $N_I$ is either increased from its
present value by up to ten times or reduced to zero, and $\chi_b$ assumes the range of values we obtained in 
our optimized parameter sets ($\chi_b = 4000 - 6500$). 
The expression in the second square bracket has a lower bound of $\approx 4.5$, and is also always positive.  We thus have:
\[
\frac{\partial g(p^*)}{\partial \alpha_1} \doteq - [``+"] - [``+"]
\]
and $g(p^*)$ is always a decreasing function of $\alpha_1$ for any dilution rate $\delta$ within our experimental
range, and for biologically reasonable choices of the experimental parameter $N_I$ and the fitted parameter $\chi_b$.

\large{(b) $g(p^*)$ is a decreasing function of $\chi_b$} \\
\normalsize 
Substituting the steady state expressions (\ref{eqs8}) into (\ref{eq20}), 
we rewrite $g$ as a function of $\chi_b$.
Taking a partial derivative of this expression with respect to $\chi_b$ yields
\be
\frac{\partial g(p^*)}{\partial\chi_b}= F(\gamma) = \frac{(\alpha_1 - 1)}{2} \gamma' \{1 - \frac{\gamma}{\sqrt{\gamma^2 + 4N_I K_c(p^*)}}\}
\label{eq24}
\ee
where $\gamma$ is defined in equation (\ref{eq15}). $F(\gamma)$ thus has the same sign as $\gamma'$, which may
be written as follows:
\[
\gamma' = \frac{\partial\gamma}{\partial\chi_b} = 
- \{ \frac{\rho_c K_b G (\delta + m + \lambda)}{\delta p [\chi_b G - (\delta + m + \lambda)]^2} \}.
\]
The expression within the curly brackets is always positive, thus $F(\gamma) < 0$ and thus
$g(p)$ is a decreasing function of $\chi_b$ for $\alpha_1 > 1$.

\large{(c) $g(p^*)$ is a decreasing function of $\delta$ for $\delta$ small,  
and an increasing function of $\delta$ when $\delta$ is large for $\alpha_1 > 1$}\\
\normalsize 
Again substituting the steady state expressions (\ref{eqs8}) into (\ref{eq20}), 
we can rewrite $g$ as a function of $\delta$. 
Taking a partial derivative of this expression with respect to $\delta$ yields
\[
\frac{\partial g}{\partial\delta}= F(\gamma)
\]
where $F(\gamma)$ is the expression on the right-hand side of equation (\ref{eq24}) and $\gamma$ is 
defined in equation (\ref{eq15}). As above, $F(\gamma)$ has the same sign as $\gamma'$, which may be written as follows:
\[
\gamma'= \frac{\partial\gamma}{\partial\delta} =\frac{\rho_C K_b}{p^*}\{ \frac{(\delta + m + \lambda)^2 - \chi_b G (m +\lambda)}{[\delta(\chi_BG- m -\lambda) -\delta^2]^2}\}
\label{eq26}
\]
This implies that $F(\gamma) < 0$ at low $\delta$, and $F(\gamma) > 0$ at high $\delta$, switching sign
at  dilution rate 
$$\delta_{\textrm{crit}} = -(m +\lambda) + \sqrt{\chi_b G(m +\lambda)}. $$ 
\pagebreak


\begin{thebibliography}
\large

\bibitem{Abrams2000}
Abrams, P. A. (2000). The evolution of predator-prey interactions: theory and evidence. 
Annual Review of Ecology and Systematics 31:79-105.

\bibitem{AbramsVos}
Abrams, P. and M. Vos. Adaptation, density dependence, and the responses of trophic level abundances to 
mortality. Evolutionary Ecology Research, \textit{in press}. 

\bibitem{Ashley2003}
Ashley, M.V., M.F. Willson, O.R.W. Pergams, D.J. O'Dowd, S.M. Gende, and J.S. Brown (2003).
Evolutionarily enlightened management. Biological Conservation 111: 115-123. 

\bibitem{Cousyn01}
Cousyn, C., L. De Meester, J.K. Colbourne, L. Brendonck, D. Verschuren,  and F. Volckaert (2001). 
Rapid, local adaptation of zooplankton behavior to changes in predation pressure in the absence of 
neutral genetic changes. Proc. Natl. Acad. Sci. USA 98: 6256-6260.

\bibitem{Cushing_etal}
Cushing, J. M.,  R. F. Costantino, B. Dennis, R. A. Desharnais, S. M. Henson (2002).
Chaos in ecology: experimental nonlinear dynamics.
Theoretical Ecology Series Volume I, Academic Press, San Diego.

\bibitem{Diek97}
Dieckmann, U. (1997). Can adaptive dynamics invade? Trends in Ecology and Evolution
12: 128-131. 

\bibitem{Dercole03}
Dercole F, J.O. Irisson, and S. Rinaldi (2003). Bifurcation analysis of a prey-predator coevolution model.
SIAM Journal on Applied Mathematics 63: 1378-1391. 

\bibitem{Ellner94}
Ellner S. P. and N. G. Hairston, Jr. (1994).
Role of overlapping generations in maintaining genetic variation in a fluctuating environment.
\textit{The American Naturalist} (143) 403--417.

\bibitem{Fussmann}
Fussmann G. F., S. P. Ellner, K. W. Shertzer, N. G. Hairston, Jr. (2000). Crossing the Hopf bifurcation in a live 
predator-prey system. \emph{ Science} (290)1358--1360.

\bibitem{Fussmann03}
Fussmann, G. F., S.P. Ellner, and N.G. Hairston, Jr. Evolution as a critical component of 
plankton dynamics. Proceedings of the Royal Society of London Series B 270: 1015-1022. 



\bibitem{Grant2002}
Grant, P.R. and B.R Grant (2002) Unpredictable evolution in a 30-year study of Darwin's finches.
Science 296, 707–-711. 

\bibitem{Hairston99}
Hairston, N.G., W. Lampert,  C.E. Cáceres, C.L. Holtmeier, L.J. Weider, U. Gaedke, J.M. Fischer, 
J.A. Fox, J.A. and D. M. Post (1999).  Lake ecosystems: Rapid evolution revealed by dormant eggs. Nature 401: 446.

\bibitem{JATREE}
Johnson, M.T.J. and A. Agrawal (2003). The ecological play of predator-prey dynamics in an
evolutionary theatre. Trends in Ecology and Evolution 18: 549--551. 

\bibitem{KK97}
Khibnik, A.I. and A.S. Kondrashov (1997). Three mechanisms of Red Queen dynamics
Proceedings of the Royal Society of London Series B 
264: 1049-1056. 

\bibitem{LeGalliard}
Le Galliard J.F., R. Ferriere, and U. Dieckmann (2003). The adaptive dynamics of altruism in spatially 
heterogeneous populations.  Evolution 57: 1-17. 

\bibitem{LawGrey89}
Law, R. and D.R. Grey (1989). Evolution of yields from populations with
age-specific cropping. Evolutionary Ecology 3: 343--359. 

\bibitem{LevinMullerLandau00}
Levin, S.A. and H.C. Muller-Landau (2000). The evolution of dispersal and seed size in
plant communities. Evolutionary Ecology Research 2: 409--435. 

\bibitem{MarrowAD}
Marrow, P., U. Dieckmann and R. Law  (1996). 
Evolutionary dynamics of predator-prey systems: An ecological perspective.
Journal of Mathematical Biology 34: 556-578. 

\bibitem{Morin99}
Morin, P. (1999). Community Ecology. Blackwell Science.

\bibitem{MuellerJoshi2000}
Mueller, L. D. and A. Joshi (2000). Stability in Model Populations. Monographs in
Population Biology (31), Princeton University Press, Princeton and Oxford.

\bibitem{pickettheaps}
Pickett-Heaps, J. D. 1975. Green Algae: Structure, Reproduction and Evolution in Selected Genera. Sinauer Associates, Sunderland MA.

\bibitem{Resnick97}
Reznick D.N., F. H. Shaw, F. H. Rodd, R. G. Shaw  (1997). 
Evaluation of the Rate of Evolution in Natural Populations of Guppies (Poecilia reticulata).
Science 275, 1934-1937.  

\bibitem{Resnick01}
Reznick, D.N. and  C. K. Ghalambor (2001). The population ecology of contemporary adaptations: what empirical 
studies reveal about the conditions that promote adaptive evolution. Genetica 112-113: 188-198.

\bibitem{Shertzer}
Shertzer K. W., S. P. Ellner, G. F. Fussmann, N. G. Hairston, Jr. (2002). Predator--prey cycles in an aquatic microcosm: testing hypotheses of mechanism.
\textit{Journal of Animal Ecology} (71) 802--815.

\bibitem{Sinervo}
Sinervo, B. et al. (2000). Density cycles and an offspring quantity and quality game driven by
natural selection. Nature 406, 985-–988. 

\bibitem{Stockwell03}
Stockwell C.A., A.P. Hendry, and M.T. Kinnison (2003). Contemporary evolution meets conservation biology.
Trends in Ecology and Evolution 18: 94--101. 

\bibitem{Thompson} 
Thompson, J.N. (1998). Rapid evolution as an ecological process. Trends in Ecology and Evolution 13: 329--332.

\bibitem{Yoshida}
Yoshida T., L. E. Jones, S. P. Ellner, G. F. Fussmann, N. G. Hairston, Jr. (2003).
Rapid evolution drives ecological dynamics in a predator-prey system.
\emph{Nature} (424) 303--306.

\bibitem{Zimmer03}
Zimmer, C. (2003) Rapid evolution can foil even the best-laid plans.
Science 300: 895. 

\end{thebibliography}
\end{document}